\documentclass{article}

\usepackage{authblk}                    
\usepackage{graphicx,epsfig,epstopdf}   
\usepackage{booktabs}                   
\usepackage{amsmath,amsthm,amssymb}     
\usepackage{color}                      
\usepackage{cite}                       

\DeclareUnicodeCharacter{2212}{-}

\graphicspath{ {figures/} }

\usepackage[paperwidth = 17cm,          
paperheight = 24cm,
left = 1.75cm,
right = 1.75cm,
top = 2cm,
bottom = 2cm]{geometry}

\usepackage[
    bookmarks         = true,
    bookmarksnumbered = true,
    bookmarksopen     = true,
    colorlinks        = true,
    linktocpage       = true,%
    urlcolor          = blue,
    linkcolor         = blue,
    citecolor         = red,
    ]{hyperref}%

\definecolor{coolblack}{rgb}{0.0, 0.18, 0.39}

\title{\textbf{Spin Trio: a dark matter scenario}}
\author[1]{Mohammad Hossein Rahimi Abkenar\thanks{rahimi.mh@fs.lu.ac.ir}}
\author[1]{Ahmad Mohamadnejad\thanks{mohamadnejad.a@lu.ac.ir}}
\author[1]{Reza Sepahvand\thanks{sepahvand.r@lu.ac.ir}}
\affil[1]{Department of Physics, Lorestan University,
Khorramabad, Iran}

\begin{document}

\maketitle

\begin{abstract}
We investigate a beyond Standard Model (SM) featuring five new fields. Four fields encompassing three distinct spin states - scalar ($ S $), spinor ($ \psi^{1,2} $), and vector ($ V_{\mu} $) - together form the multi-component dark matter (DM), while the fifth (scalar) field ($ \phi $) carries a unit charge under a dark $ U_{D}(1) $ gauge symmetry, enabling SM-DM interactions via the Higgs portal. Although the model maintains classical scale invariance, loop effects break electroweak symmetry. The parameter space is constrained by scale invariance, DM relic density, and direct detection results. Our study aims to identify feasible model regions and evaluate detectability in future experiments. We analyze processes like DM annihilations, semi-annihilations, and conversions, integrating them into Boltzmann equations to calculate DM abundances. Random parameter scans reveal regions compatible with current data, including constraints from direct detection experiments like XENONnT and PandaX-4T . Our results show the model's viability across a broad range of DM masses.
\end{abstract}


\numberwithin{equation}{section}

\section{Introduction} \label{sec1}
According to some evidence such as galaxy rotation curves~\cite{Rubin:1970zza}, CMB anisotropy~\cite{WMAP:2003elm}, gravitational lensing~\cite{Bartelmann:1999yn} and radial velocities of the galaxies in a cluster~\cite{Zwicky:1937zza}, the Universe contains significantly more mass than what is observed. Cosmic evidence indicates that about 27\% of the Universe is DM, approximately 5\% is ordinary (baryonic) matter, and roughly 68\% is dark energy~\cite{Planck:2018vyg}. These percentages emphasize the crucial role dark matter plays in the Universe's structure and dynamics. As a major component of the total mass content, yet remaining undetectable through direct observation, dark matter represents one of the key unresolved issues in astro-particle physics today. Many of its particle properties, including spin, mass, and interactions, are still not understood.

Recent theoretical advances have led physicists to investigate models beyond the SM to explain the particle nature of DM, especially in light of new experimental data and constraints from direct detection experiments~\cite{Bertone:2004pz,Cirelli:2024ssz}. One viable possibility is that DM comprises multiple weakly interacting massive particles (WIMPs)~\cite{Roszkowski:2017nbc,Arcadi:2017kky}, each contributing to the observed DM density~\cite{Zurek:2008qg,Profumo:2009tb,Aoki:2012ub,Biswas:2013nn,Gu:2013iy,Aoki:2013gzs,Kajiyama:2013rla,Bian:2013wna,Bhattacharya:2013hva,Geng:2013nda,Esch:2014jpa,Dienes:2014via,Bian:2014cja,Geng:2014dea,DiFranzo:2016uzc,Aoki:2016glu,DuttaBanik:2016jzv,Pandey:2017quk,Borah:2017xgm,Herrero-Garcia:2017vrl,Ahmed:2017dbb,PeymanZakeri:2018zaa,Aoki:2018gjf,Chakraborti:2018lso,Bernal:2018aon,Poulin:2018kap,Herrero-Garcia:2018qnz,YaserAyazi:2018lrv,Elahi:2019jeo,Borah:2019epq,Borah:2019aeq,Bhattacharya:2019fgs,Biswas:2019ygr,Nanda:2019nqy,Yaguna:2019cvp,Belanger:2020hyh,VanDong:2020bkg,Khalil:2020syr,DuttaBanik:2020jrj,Hernandez-Sanchez:2020aop,Chakrabarty:2021kmr,Yaguna:2021vhb,DiazSaez:2021pmg,DiazSaez:2021pfw,Mohamadnejad:2021tke,Yaguna:2021rds,Ho:2022erb,Kim:2022sfg,Das:2022oyx,Belanger:2022esk,Hosseini:2023qwu}. A natural motivation for these multi-component DM scenarios arises from the SM itself, which includes various fields with different spins.

This paper presents a model featuring three key motives. It goes beyond the SM to address the DM problem, the hierarchy problem, and vacuum instability. To tackle the hierarchy problem, the model incorporates classical conformal symmetry~\cite{Bardeen:1995kv}. In contrast, the SM encounters issues with vacuum instability~\cite{Degrassi:2012ry,Tang:2013bz}, and models beyond the SM that include bosonic degrees of freedom may effectively resolve this problem. Furthermore, due to the strict constraints on direct detection for single DM, multi-component models are considered to be more suitable. Regarding these motives, we introduce a new model that incorporates five beyond SM fields. Four of them collectively constitute a DM - namely a scalar ($S$), two spinors ($\psi^{1,2}$), and a vector ($V_{\mu}$) - each characterized by distinct spins. The incorporation of a fifth scalar field ($\phi$), which possesses a unit charge under a dark $U_{D}(1)$ gauge symmetry, enables interactions between the SM and DM through the Higgs portal. This interaction is pivotal for bridging the gap between theoretical constructs and observable phenomena, allowing us to test DM candidates within experimental frameworks. The model maintains classical scale invariance; however, electroweak symmetry breaking occurs due to loop effects. In this framework, all particle masses are dynamically generated through the Coleman-Weinberg mechanism~\cite{Coleman:1973jx}. By exploring the expanded parameter space of the model, we aim to identify feasible regions where our model can adequately account for DM relic density and comply with the latest constraints from direct detection experiments. Understanding these parameters is essential for delineating the conditions under which our model can be validated, as well as forecasting its potential observable consequences.

To assess the detectability of our model in future experiments, we carefully analyze processes like annihilations, semi-annihilations, and conversions that influence DM abundances. These processes are integrated into Boltzmann equations, which allow for the calculation of DM relic densities using tools like {\tt micrOMEGAs}~\cite{Alguero:2023zol}. By employing random parameter scans, we can isolate regions of parameter space that align with current experimental evidence, notably incorporating constraints from experiments like XENONnT~\cite{XENON:2023cxc} and PandaX-4T ~\cite{PandaX-4T:2021bab}. These experiments are progressively nearing what is known as the neutrino floor, which represents the maximum sensitivity achievable by future direct detection experiments~\cite{Billard:2013qya}. Our findings indicate the model's viability across a broad spectrum of DM masses, ultimately presenting a comprehensive framework that could lead to promising signals in future direct detection experiments. This exploration not only enhances our understanding of DM but also sheds light on the potential for new physics beyond the SM, contributing to the ongoing quest to decipher the nature of our Universe's unseen constituents.

The paper is organized as follows: Section~\ref{sec2} introduces the model, while section~\ref{sec3} and section~\ref{sec4} describe its DM phenomenology, i.e., DM relic density and direct detection, respectively. Section~\ref{sec5} presents and analyzes the main results from various parameter space scans. Finally, section~\ref{sec6} concludes the discussion.

\section{The model} \label{sec2}
We propose a model comprising a (real) scalar field \( S \), two spinor fields \( \psi^{1,2} \) (including right-handed \( \psi^{1,2}_R \) and left-handed \( \psi^{1,2}_L \)), and a vector field \( V_\mu \) as DM, along with a (complex) scalar field \( \phi \) serving as an intermediate particle.
All of these new fields are singlet under SM gauge group. We implement a discrete symmetry under which, the new fields transform as follows
\begin{equation} \label{z2symmetry}
\phi \rightarrow \phi^{*}, \quad S \rightarrow -S, \, \quad V_{\mu} \rightarrow - V_{\mu}, \, \quad \psi_L^1 \rightarrow -\psi_L^2  \quad \text{and}  \, \quad \psi_R^1 \rightarrow -\psi_R^2,
\end{equation}
while all SM particles are even. In our setup, the scalar field \( \phi \) and spinor fields \( (\psi^{1,2}_R \) and \( \psi^{1,2}_L) \) carry charge under a dark \( U_D(1) \) gauge symmetry, with the vector field \( V_{\mu} \) serving as the gauge field (see Table~\ref{charge}). All SM particles are singlets under the dark gauge symmetry, while the new fields transform as
\begin{align} \label{invariant}
& S \rightarrow e^{i Q_S \alpha(x)} S, \nonumber \\
& \psi_L^a \rightarrow e^{i Q_L^a \alpha(x)} \psi_L^a, \nonumber \\
& \psi_R^a \rightarrow e^{i Q_R^a \alpha(x)} \psi_R^a, \nonumber \\
& \phi \rightarrow e^{i Q_{\phi} \alpha(x)} \phi, \nonumber \\
& V_{\mu} \rightarrow V_{\mu} - \frac{1}{g_v} \partial_{\mu}{\alpha(x)} ,
\end{align}
where $ a \in \{1, 2\} $.

\begin{table}[htb] 
\centering
\parbox{6.5cm}{\caption{The charges associated with the particles in the dark sector with respect to the newly introduced \( U_D(1) \) symmetry.}\label{charge}}

\vspace{7pt}
\begin{tabular}{ l  l  l  l  l  l  l  l}
\hline
field&$\phi$&$S$&$V_{\mu}$&$\psi_L^1$&$\psi_R^1$&$\psi_L^2$&$\psi_R^2$ \vspace{2pt} \\
\hline
dark charge ($ Q $)&$1$&0&0&$\frac{1}{2}$&$-\frac{1}{2}$&$-\frac{1}{2}$&$\frac{1}{2}$ \vspace{2pt} \\
\hline
\end{tabular}
\end{table}

Regarding renormalizable interactions, the most general Lagrangian is given by:
\begin{align}  \label{tlagrangian}
{\cal L} ={\cal L}_{SM}+\frac{1}{2}(\partial_{\mu} S)(\partial^{\mu} S) + (D_{\mu} \phi)^{*} (D^{\mu} \phi) -\frac{1}{4} V_{\mu \nu} V^{\mu \nu}- V(H,S,\phi) \nonumber \\
+\sum_{a=1}^2 \left(  i\bar\psi_L^a  \gamma^{\mu}D_{\mu}\psi_L^a+ i \bar\psi_R^a \gamma^{\mu}D_{\mu}\psi_R^a\right)
-g_{\phi,1} \phi \bar\psi_L^1 \psi_R^1 - g_{\phi,2} \phi^{*} \bar\psi_L^2 \psi_R^2 + {\text{H.C.}} ,
\end{align}
where $ {\cal L} _{SM} $, $D_{\mu}= (\partial_{\mu} + i Q g_{v} V_{\mu})$ and $V_{\mu \nu}= \partial_{\mu} V_{\nu} - \partial_{\nu} V_{\mu} $ are the SM Lagrangian without the Higgs potential term, covariant derivative, and field strength tensor, respectively. 
Due to the symmetry condition (\ref{z2symmetry}), it follows that \( g_{\phi,1}=g_{\phi,2}=g_{\phi} \), which consequently implies \( M_{\psi_1}=M_{\psi_2}=M_{\psi} \).
The Lagrangian (\ref{tlagrangian}) is invariant under both discrete (\ref{z2symmetry}), and local gauge transformations (\ref{invariant}).
Finally, the most general renormalizable scale/gauge invariant potential $V(H,\phi,S)$ is
\begin{align} \label{vlagrangian}
V(H,\phi,S) = & \, \lambda_{H} (H^{\dagger}H)^{2} + \lambda_{\phi} (\phi^{*}\phi)^{2}  
+ \lambda_{H \phi} (H^{\dagger}H)(\phi^{*}\phi) \nonumber \\
&+ \frac{1}{2} \lambda_{H S} (H^{\dagger}H)S^2
+ \frac{1}{2} \lambda_{\phi S} (\phi^{*}\phi)S^2 + \frac{1}{4} \lambda_{S} S^4.
\end{align}

To determine if the model is anomaly-free, we need to analyze potential gauge anomalies, which occur when classical symmetries fail at the quantum level and can lead to inconsistencies in the theory. Gauge anomalies arise from triangle diagrams involving three gauge bosons, requiring that the contributions from all fermions cancel for anomaly freedom. In our model, the dark \( U_D(1) \) symmetry introduces a new gauge group, necessitating the examination of the cubic anomaly \( [U_D(1)]^3 \) from \( U_D(1) \)-charged fermions. 
The contributions to the \( [U_D(1)]^3 \) anomaly are 
\begin{equation} \label{anomaly1}
\text{Anomaly} \propto \sum_{a=1}^2 (Q_L^a)^3 - \sum_{a=1}^2 (Q_R^a)^3.
\end{equation}
Substituting the charges from Table \ref{charge}, we find:
\(
\left(\frac{1}{2}\right)^3 + \left(-\frac{1}{2}\right)^3 - \left(-\frac{1}{2}\right)^3 - \left(\frac{1}{2}\right)^3 = 0.
\)
Thus, the \( [U_D(1)]^3 \) anomaly cancels. Regarding mixed anomalies, the SM fermions do not contribute as they are singlets under \( U_D(1) \), and similarly, the new fermions \( \psi^{1,2} \) are singlets under the SM gauge group, resulting in no mixed anomalies.

In this model, DM components associated with scalar and spinor fields, $ S $ and $ \psi^a $, are stable. In addition, if $ M_V < 2 M_{\psi} $, then all scalar, spinor, and vector components are viable DM candidates.
This condition regarding the mass of the spinor and vector DM is essential. The gauge vector has a decay channel \( V \rightarrow \psi \overline{\psi} \), which renders the vector DM unstable. However, under the condition \( M_V < 2 M_{\psi} \), this decay mode would violate energy conservation and therefore will not take place. Consequently, we maintain this condition throughout our paper. 
Both scalar $\phi$ and Higgs fields can receive vacuum expectation values (VEVs), breaking respectively the $U_D(1)$ and electroweak gauge symmetries. In unitary gauge, the imaginary component of $\phi$ can be absorbed as the longitudinal part of $ V_{\mu} $ making it massive after spontaneous symmetry breaking. In this gauge,
we have
\begin{equation} \label{gauge}
H = \frac{1}{\sqrt{2}} \begin{pmatrix}
0 \\ h_{1} \end{pmatrix} \, \, \, \text{and} \, \, \, \phi = \frac{1}{\sqrt{2}} h_{2},
\end{equation}
where $h_{1}$ and $h_{2}$ are (real) scalar fields which can receive VEVs. By replacing Eqs. (\ref{gauge}) in Eq. (\ref{vlagrangian}), the tree-level potential can be expressed as
\begin{equation} \label{vtree}
V^{tree} = \frac{1}{4} \lambda_{H} h_{1}^{4} + \frac{1}{4} \lambda_{\phi} h_{2}^{4} + \frac{1}{4} \lambda_{\phi H} h_{1}^{2} h_{2}^{2} + \frac{1}{4} \lambda_{H S} h_{1}^2 s^2 + \frac{1}{4} \lambda_{\phi S} h_{2}^2 s^2 + \frac{1}{4} \lambda_{s} s^4  . 
\end{equation}

Local minimum of $ V^{tree} $, corresponds to VEVs $ \langle h_{1} \rangle = \nu_{1} $ and $ \langle h_{2} \rangle = \nu_{2} $. 
The first necessary condition to calculate the minimum of $V^{tree}$ is:
\begin{align} 
& \frac{\partial V^{tree}}{\partial h_{i}} \bigg\rvert_{\nu_{1},\nu_{2}} = 0, \label{hfirstcondition} \\
&  \frac{\partial V^{tree}}{\partial s} = 0 . \label{sfirstcondition}
\end{align}
Imposing discrete symmetry, $ S \rightarrow -S $, even after symmetry breaking, makes scalar field $ S $ stable and a viable DM candidate. Therefore, among $ h_{1} $, $ h_{2} $ and $ S $, only $ h_{1} $ and $ h_{2}$ can receive VEVs and we assume $ \langle S \rangle = 0 $. Now Eq. (\ref{hfirstcondition}) gives
\begin{align} 
& \frac{\nu_1^{2}}{\nu_2^{2}}= - \frac{\lambda_{\phi H}}{2 \lambda_H}, \label{nu1} \\
& \frac{\nu_2^{2}}{\nu_1^{2}}= - \frac{\lambda_{\phi H}}{2 \lambda_\phi} . \label{nu2}
\end{align}
By solving Eqs. (\ref{nu1}) and (\ref{nu2}) we obtain the following equation
\begin{equation} \label{lafh}
\lambda_{\phi H}^2=4 \lambda_H \lambda_\phi .
\end{equation}
The second necessary condition for $V^{tree}$ to have a minimum is
\begin{align} \label{secondcondition}
 \frac{\partial^{2} V^{tree}}{\partial h_{i} ^{2}} \bigg\rvert_{\nu_{1},\nu_{2}} > 0 .
\end{align}
By solving condition (\ref{secondcondition}) along with Eqs. (\ref{nu1}) and (\ref{nu2}) we obtain the following new constraints:
\begin{equation} \label{signs}
\lambda_{\phi H} < 0 \;\; , \;\; \lambda_H > 0 \;\; , \;\; \text{and} \;\;\; \lambda_\phi > 0 .
\end{equation}
Now considering the Hessian matrix defined as
\begin{equation} \label{Hessianmatrix}
H_{ij} (h_{1},h_{2}) \equiv \frac{\partial^{2} V^{tree}}{\partial h_{i} \partial h_{j}} ,
\end{equation}
the third and last necessary condition for $V^{tree}$ to have a minimum is
\begin{equation} \label{determinant}
det(H (\nu_{1},\nu_{2})) > 0 ,
\end{equation}
but, condition~(\ref{determinant}) will not be satisfied, because $ det(H (\nu_{1},\nu_{2})) = 0 $.
When the determinant of the Hessian matrix equals zero, the second derivative test does not provide conclusive information, meaning the point $ (\nu_{1},\nu_{2}) $ could represent a minimum, maximum, or saddle point. However, in our scenario, constraint (\ref{nu1}) (or (\ref{nu2})) establishes a direction, referred to as a flat direction, where $ V^{tree} = 0 $. This indicates a stationary line or a line of local minima.

Except the flat direction, where $ V^{tree} = 0 $, the potential is positive. Consequently, the overall potential of the theory will be primarily influenced by higher-loop contributions along the flat direction, particularly by the one-loop effective potential. Incorporating the one-loop effective potential, \( V_{eff}^{1-loop} \), can introduce a slight curvature in the flat direction, identifying a specific value along the ray as the minimum, where \( V_{eff}^{1-loop} < 0 \), and the vacuum expectation value is \( \nu^{2} = \nu_{1}^{2} + \nu_{2}^{2} \), characterized by a renormalization group scale \( \Lambda \). At the minimum of the one-loop effective potential, with \( V^{tree} \geqslant 0 \) and \( V_{eff}^{1-loop} < 0 \), the minimum of \( V_{eff}^{1-loop} \) along the flat direction (where \( V^{tree} = 0 \)) constitutes a global minimum of the full potential. Thus, spontaneous symmetry breaking occurs, and we should replace \( h_{1} \) with \( \nu_{1} + h_{1} \) and \( h_{2} \) with \( \nu_{2} + h_{2} \). This process breaks the electroweak symmetry, resulting in a vacuum expectation value of \( \nu_{1} = 246 \) GeV. We will first examine the tree-level potential. As \( h_{1} \) and \( h_{2} \) mix with one another, they can be expressed in terms of the mass eigenstates \( H_{1} \) and \( H_{2} \) as follows
\begin{equation} \label{exchangematrix}
\begin{bmatrix}
H_1 \\ 
H_2
\end{bmatrix}=\begin{bmatrix}
\cos \alpha & - \sin \alpha \\
\sin \alpha & \;\;\; \cos \alpha
\end{bmatrix} \begin{bmatrix}
h_1 \\ h_2
\end{bmatrix},
\end{equation}
where \( H_{2} \) lies along the flat direction, implying that \( M_{H_{2}} = 0 \), while \( H_{1} \) is oriented perpendicular to the flat direction and is identified as the SM-like Higgs observed at the LHC, with \( M_{H_{1}} = 125 \) GeV. Following the symmetry breaking, we establish the following constraints
\begin{align}
& \sin \alpha =  \frac{\nu_{1}}{\sqrt{\nu_{1}^{2}+\nu_{2}^{2}}}, \nonumber \\
& g_{V}=\frac{M_{V}}{\nu_{2}}, \nonumber \\
& g_{\phi}=\frac{\sqrt{2} M_{\psi}}{\nu_{2}}, \nonumber \\
& \lambda_{H} = \frac{M_{H_{1}}^{2} cos^2\alpha}{2 \nu_{1}^{2}}, \nonumber  \\
& \lambda_{\phi} = \frac{M_{H_{1}}^{2} sin^2\alpha}{2 \nu_{2}^{2}}, \nonumber  \\
& \lambda_{\phi H} =  - \frac{M_{H_{1}}^{2} \sin \alpha \cos \alpha}{ \nu_{1}\nu_{2}}, \nonumber \\
& \lambda_{H S} = \frac{2 M_{S}^{2} - \lambda_{\phi S} \nu_{2}^2}{\nu_{1}^2},
 \label{constrins}
\end{align}
where $M_{V}$, $M_{S}$, and $M_{\psi}$ are the masses of vector, scalar and spinor DM, respectively.
Finally, by using the Gildener-Weinberg formalism \cite{Gildener:1976ih} for one-loop corrections of potential, the mass of $H_2$ is determined by the following equation
\begin{equation}  \label{MH2}
M_{H_{2}}^{2} = \frac{1}{8 \pi^{2} \nu^{2}} \left( M_{H_{1}}^{4} + 6  M_{W}^{4} + 3  M_{Z}^{4} + 3  M_{V}^{4} + M_{S}^{4} - 8 M_{\psi}^{4} - 12 M_{t}^{4}   \right).
\end{equation}
The condition, $ M_{H_{2}}^{2} > 0 $, puts a constraint on the masses of DM components.

We choose free parameters of the model as follows:
\begin{equation} \label{freeparameters}
M_S \;\;,\;\; M_{\psi} \;\;,\;\; M_V \;\;,\;\; \lambda_{\phi S} \;\;,\;\; \nu_2 .
\end{equation}
Note that $\lambda_S$ is not a relevant parameter in calculations of DM relic density and DM-nucleon cross section.
For perturbativity, we also apply
\begin{align} \label{4pi}
& \!\!\!\!\!\! -4\pi < \lambda_{HS} \;,\; \lambda_{\phi S} \; \;,\; g_{\phi} \;,\; g_{V} < +4\pi \nonumber \\
& \;\;\;\;\;\;\;\;\;\;\;\;\; 0 < \lambda_H \;,\; \lambda_\phi < +4\pi \nonumber \\
& \;\;\;\;\;\;\;\;\;\;\;\;\; -4\pi < \lambda_{\phi H} < 0 .
\end{align}
In the upcoming sections, we will evaluate the consistency of our model with DM relic density and data from direct detection experiments.

\section{Relic density} \label{sec3}
\begin{figure} [htb] 
\centerline{\hspace{0cm}\epsfig{figure=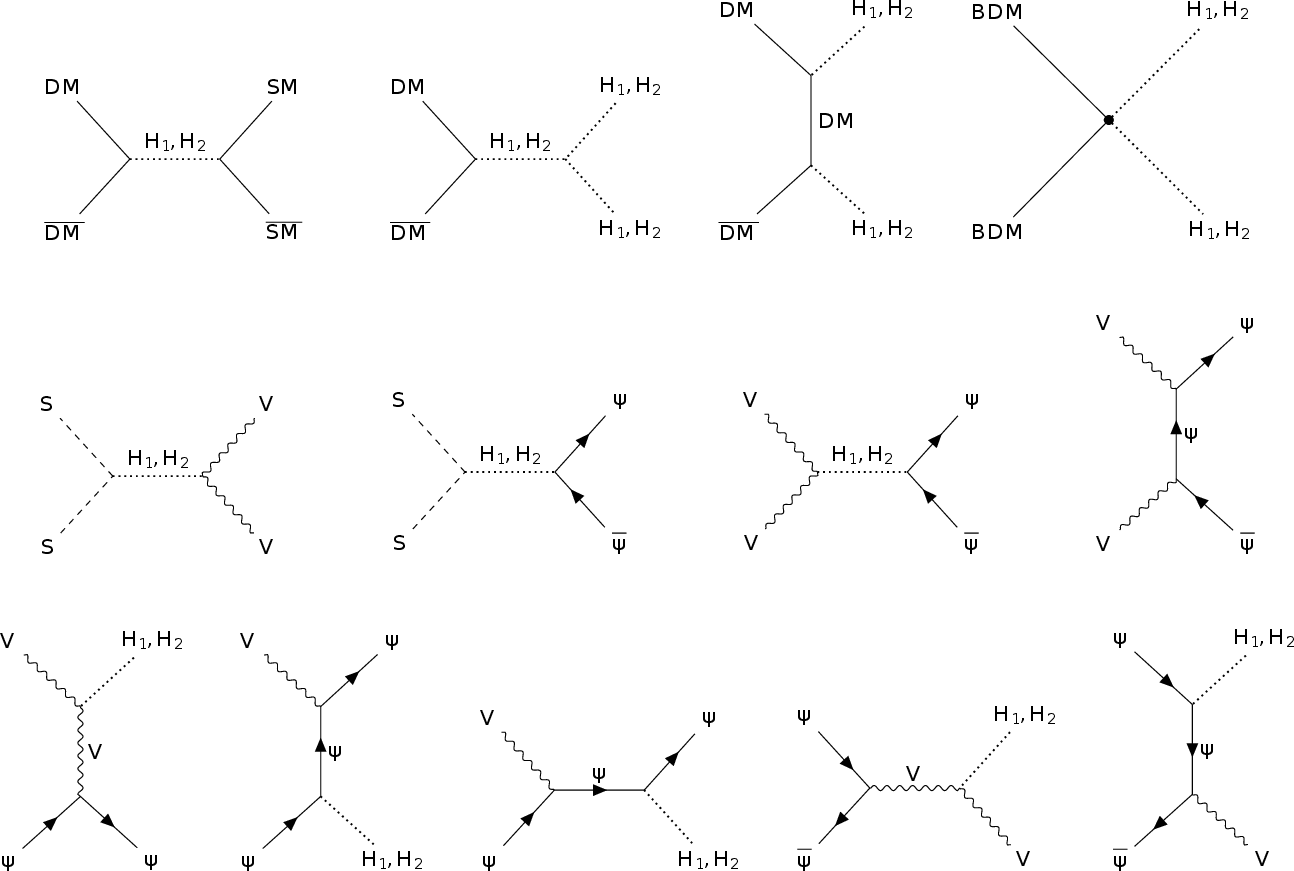,width=12.5cm}}
\caption{Relevant Feynman diagrams for relic density. line 1: DM annihilation. line 2: DM conversion. line 3: DM semi-annihilation. SM, DM, and BDM, stand for Standard Model massive fermions and gauge bosons, dark matter particles, and bosonic dark matter particles, respectively. The field \(\psi\) can correspond to either \(\psi^1\) or \(\psi^2\).} \label{FeynmanRD}
\end{figure}

Assuming only $ 2 \rightarrow 2 $ reactions, the Boltzmann equation describing the evolution of the abundance of the $ i $-th DM candidate reads~\cite{Alguero:2023zol}
\begin{equation} \label{Boltzmann1}
\frac{dn_i}{dt} = - \sum_{\alpha \leq \beta ; \; \gamma \leq \delta} C_{\alpha \beta} n_{\alpha} n_{\beta} \langle v \sigma_{\alpha \beta \rightarrow \gamma \delta} \rangle (\delta_{i \alpha} + \delta_{i \beta} - \delta_{i \gamma} - \delta_{i \delta})-3H(T)n_i,
\end{equation}
where $ n_i $ with $ i=1, 2, 3 $ is the number density for each DM species, $ C_{\alpha \beta} $ is a combinatoric factor, $ C_{\alpha \beta} = 1/2 $ if $ \alpha = \beta $ and $ 1 $ otherwise, and $ H(T) $ is the Hubble expansion rate. Greek indices run over $ 0, 1, 2, 3 $, with zero refering to the sector that includes all SM particles as well as any new particle types that share the same discrete symmetry transformation characteristics as the SM, i.e., $ H_2 $ particle.
Considering the conservation of entropy, $ \frac{ds}{dt} = −3Hs $, we get
\begin{equation} \label{Boltzmann2}
3H \frac{dY_{i}}{ds} = \sum_{\alpha \leq \beta ; \; \gamma \leq \delta} C_{\alpha \beta} Y_{\alpha}Y_{\beta} \langle v \sigma_{\alpha \beta \rightarrow \gamma \delta} \rangle (\delta_{i \alpha} + \delta_{i \beta} - \delta_{i \gamma} - \delta_{i \delta}),
\end{equation}
where $ Y_i=n_i/s $. By solving these Boltzmann equations, we can determine the relic densities, $\Omega_{S,\psi,V}$. In our model, the DM fermions $ \psi^{1,2} $ possess identical masses and contribute equally to the relic abundance of fermionic DM ($ \psi $).
We have included the impact of the various processes—annihilations, semi-annihilations, and conversions—that influence the relic density of DM components (see figure \ref{FeynmanRD}).

For the numerical solution of this system of coupled equations, we utilize the last version of {\tt micrOMEGAs}~\cite{Alguero:2023zol} via
{\tt LanHEP}~\cite{Semenov:2014rea}. In this new version, the code has been updated to support models with three DM particles.  

Given that we are examining a multi-component DM scenario, we need to compare the total relic densities of each DM particle, $\Omega_S + \Omega_{\psi} + \Omega_V$, with the observed DM abundance, $\Omega_{\text{DM}}$. The value of $\Omega_{\text{DM}}$, as reported by the PLANCK collaboration \cite{Planck:2018vyg}, carries an estimated theoretical uncertainty around 10 percent,
\begin{align} \label{0.12}
\Omega_{\text{DM}}h^2=[0.11,0.13]. 
\end{align}
As will be demonstrated, this DM constraint significantly limits the viable parameter space of the model.

\begin{figure} [!htb] 
\centerline{\hspace{0cm}\epsfig{figure=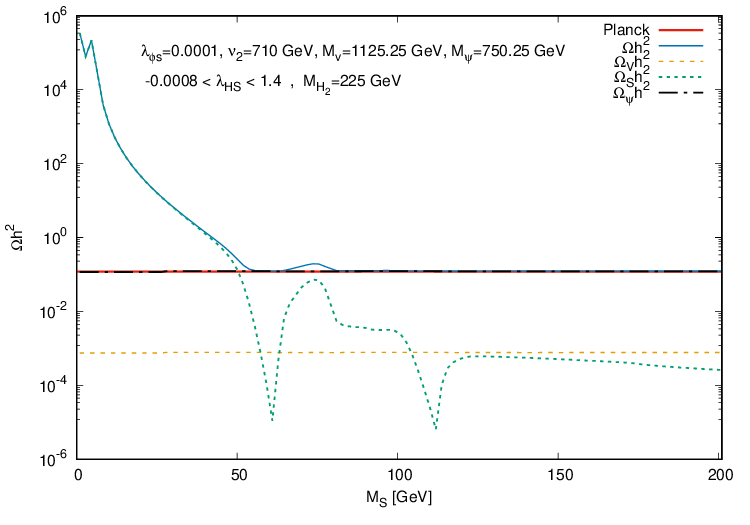,width=7.5cm}}
\caption{Relic densities of DM components as a function of the scalar DM mass.} \label{omega-sc}
\end{figure}

\begin{figure} [!htb] 
\centerline{\hspace{0cm}\epsfig{figure=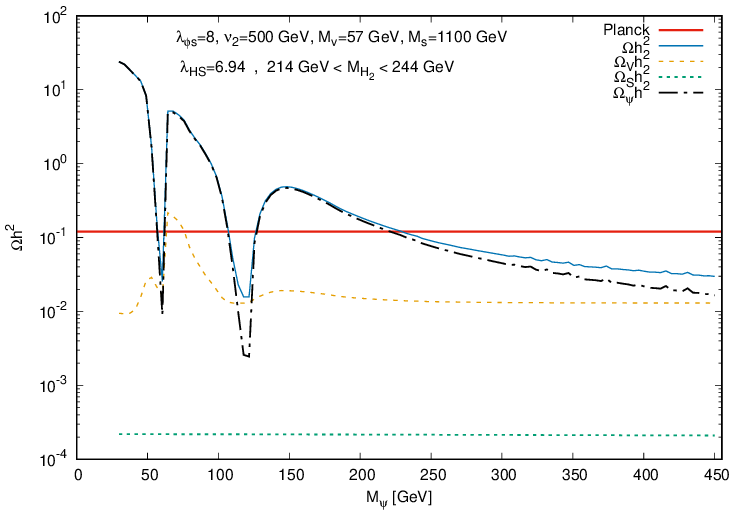,width=7.5cm}}
\caption{Relic densities of DM components as a function of the spinor DM mass.} \label{omega-psi}
\end{figure}

\begin{figure} [!htb] 
\centerline{\hspace{0cm}\epsfig{figure=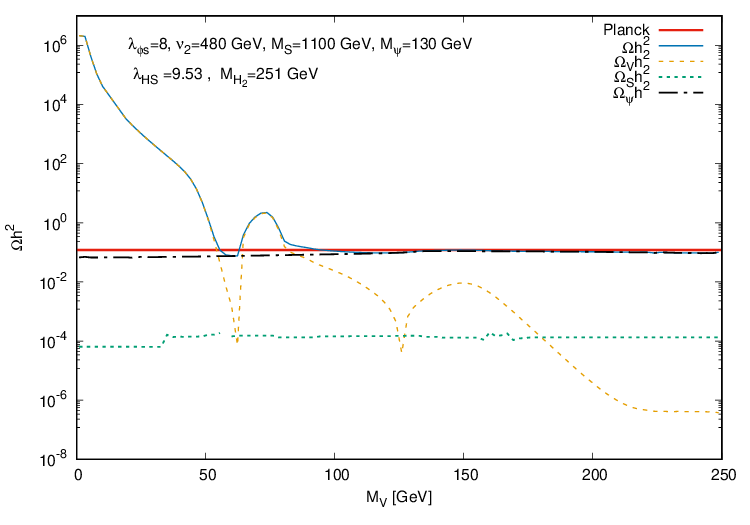,width=7.5cm}}
\caption{Relic densities of DM components as a function of the vector DM mass.} \label{omega-v}
\end{figure}

In the rest of this section, we will depict DM relic density diagrams versus independent parameters outlined in (\ref{freeparameters}).
When one DM component is more massive than the other two, its conversion into the lighter components becomes significant. Consequently, we anticipate that the primary contribution to the relic density of DM comes from the lightest species. Conversely, the freeze-out temperature for the lightest species, approximately given by \( T_{fo} \simeq M_{DM}/25 \), is lower than that of the heavier components. This indicates that the freeze-out of the lightest DM components occurs after that of the heavier ones, leading to a lower abundance for the lighter particles. These two effects are in competition with one another. 
Another notable characteristic is the dual reduction of relic density occurring at the \( H_1 \) and \( H_2 \) resonance points (specifically at \( M_{DM}\simeq\frac{M_{H_1}}{2}=62.5 \) and \( M_{DM}\simeq\frac{M_{H_2}}{2} \)). This is evident in figures \ref{omega-sc} through \ref{omega-v}.

In figure~\ref{omega-sc}, the relic densities of vector and spinor DM  does not vary dramatically with changes in the mass of scalar DM, as anticipated. The freeze-out temperatures for spinor and vector DM are almost equal and higher than that of scalar DM. Note that the relic density of spinor DM is greater than that of vector DM, as vector DM has a greater mass. This is due to the fact that heavier DM can convert into lighter one, while the opposite conversion is less likely.
The scalar DM relic density shows two decreases and one increase. The decreases happen at the \( H_1 \) and \( H_2 \) resonance points, whereas the increase occurs at 
\begin{equation} \label{lambdaHS}
M_{S} = \nu_2 \sqrt{\frac{\lambda_{\phi S}}{2}} \simeq  5 \, \, \text{GeV} \quad \Rightarrow \quad   \lambda_{H S} = \frac{2 M_{S}^{2} - \lambda_{\phi S} \nu_{2}^2}{\nu_{1}^2}=0,
\end{equation}
where a decline in scalar DM annihilation results in a rise in scalar DM relic density.

In figure~\ref{omega-psi}, there are two minima observed in the spinor DM relic density around 62 GeV and 120 GeV. These reductions in relic density at these points are associated with the resonances of \( H_1 \) and \( H_2 \), as mentioned earlier.
In figure \ref{omega-v}, at first, \( M_V \) is less than \( M_{\psi} \), resulting in \( \Omega h^2 \) being determined by \( \Omega_V h^2 \). However, as \( M_V \) increases, spinor DM becomes lighter than the vector DM, leading \( \Omega h^2 \) to match \( \Omega_{\psi} h^2 \). As before, two minima are evident in the vector DM relic density at approximately 62 GeV and 120 GeV, corresponding to the \( H_1 \) and \( H_2 \) resonance points. The eventual drop in \( \Omega_V h^2 \) as \( M_V \) increases is due to the rise of \( g_{V} = M_{V} / \nu_{2} \) with increasing \( M_V \), which enhances the annihilation of vector DM and ultimately results in a decrease in \( \Omega_V h^2 \).

\begin{figure} [!htb] 
\begin{center} 
\centerline{\hspace{0cm}\epsfig{figure=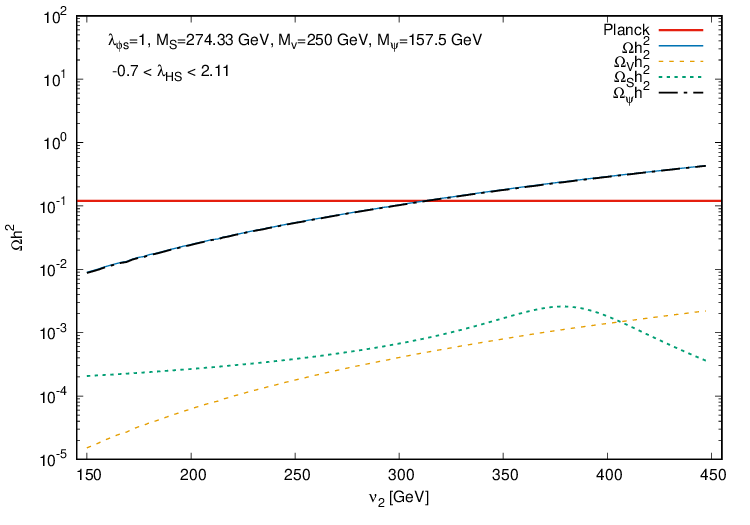,width=7.5cm}}
\centerline{\vspace{-0.7cm}} 
\caption{Relic densities of DM components as a function of $\nu_2$.} \label{omega-nu2}
\end{center} 
\end{figure}

\begin{figure}[!htb] 
\begin{center}
\centerline{\hspace{0cm}\epsfig{figure=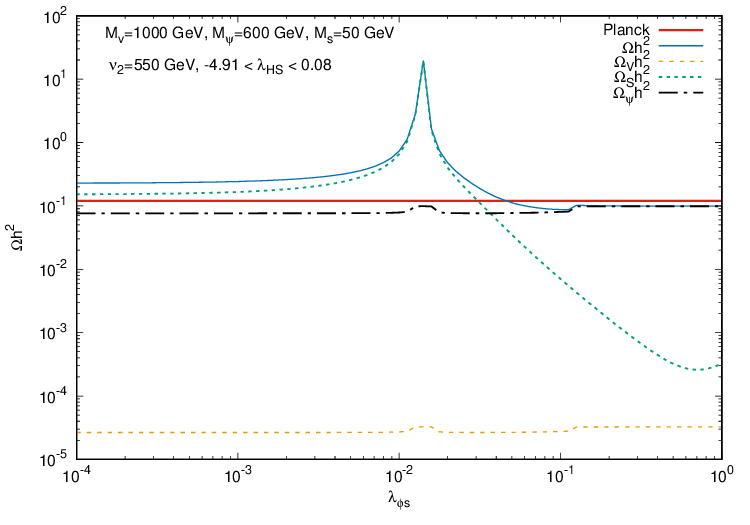,width=6.5cm}\hspace{0cm}\epsfig{figure=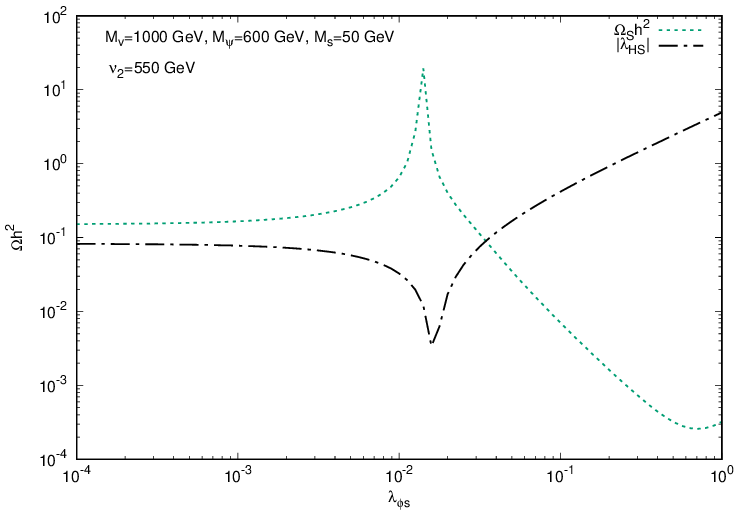,width=6.5cm}}
\centerline{\vspace{-0.7cm}}
\caption{Left: relic densities of DM components as a function of the coupling $\lambda_{\phi S}$. Right: scalar DM relic density and $|\lambda_{H S}|$ versus $\lambda_{\phi S}$.} \label{omega-lafs}
\end{center} 
\end{figure} 

\begin{figure} [!htb] 
\begin{center} 
\centerline{\hspace{0cm}\epsfig{figure=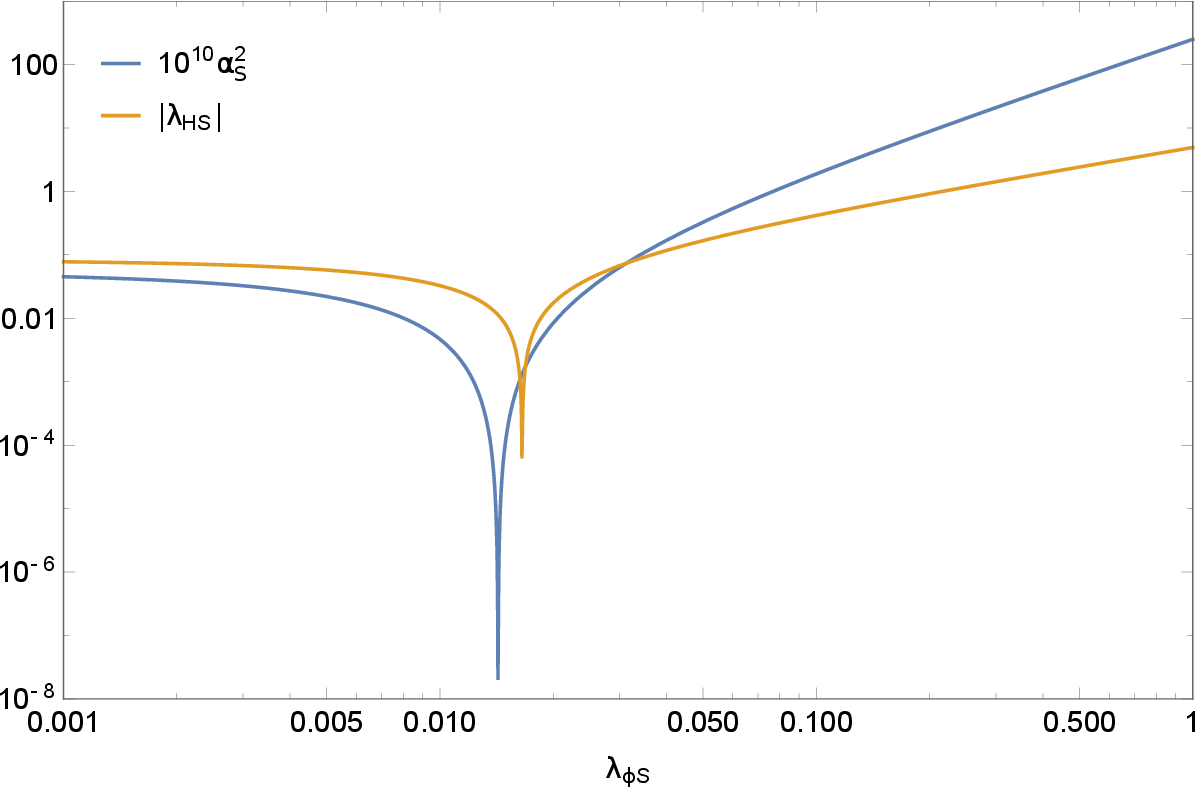,width=7.5cm}}
\centerline{\vspace{-0.7cm}} 
\caption{$|\lambda_{H S}|$ and $ \alpha_S^2 $ versus $\lambda_{\phi S}$ with the same parameters as figure~\ref{omega-lafs}.} \label{lafsii}
\end{center} 
\end{figure}

In figure~\ref{omega-nu2}, both \( \Omega_V h^2 \) and \( \Omega_{\psi} h^2 \) are ascending. This is due to the fact that \( g_{V} = M_{V} / \nu_{2} \) and \( g_{\phi} = \sqrt{2} M_{\psi} / \nu_{2} \) decrease with increasing \( \nu_2 \), leading to a reduction in the annihilation of vector and spinor DM and an increase in their relic densities. In contrast, for scalar DM, the relic density initially rises until 
\begin{equation} \label{lambdaHSnu2}
\nu_2  = M_{S} \sqrt{\frac{2}{\lambda_{\phi S}}} \simeq  384 \, \, GeV \quad \Rightarrow \quad   \lambda_{H S} = \frac{2 M_{S}^{2} - \lambda_{\phi S} \nu_{2}^2}{\nu_{1}^2}=0,
\end{equation}
and then falls again.

The annihilation rate of scalar DM depends on the couplings $\lambda_{\phi S}$ and $\lambda_{H S}$, while vector and spinor DM annihilation rates are largely unaffected by these parameters, leading to constant relic densities despite variations in $\lambda_{\phi S}$ and $\lambda_{H S}$ (refer to figure~\ref{omega-lafs} (left)). In figure~\ref{omega-lafs} (right), the values of $\lambda_{\phi S}$ and $\lambda_{H S}$ are interdependent, as indicated by the final equation in (\ref{constrins}). At the point where $\lambda_{\phi S} = 2 M_S^2 / \nu_2^2 \simeq 0.016$, we find that $\lambda_{H S} \simeq 0$. When $\lambda_{\phi S}$ is significantly less than 0.016, $\lambda_{H S}$ remains approximately constant at $2 M_S^2 / \nu_1^2 \simeq 0.08$, which is much larger than $\lambda_{\phi S}$, making it the dominant coupling in scalar DM annihilation. Consequently, for $\lambda_{\phi S} \ll 0.016$, the relic density of scalar DM also remains constant. As $\lambda_{\phi S}$ increases and becomes comparable to $\lambda_{H S}$, both couplings must be considered together. The effective coupling for the annihilation of scalar DM into fermions is represented by $\alpha_S$ (see~(\ref{couplings})), which is dependent on both $\lambda_{\phi S}$ and $\lambda_{H S}$. Figure~\ref{lafsii} illustrates both $|\lambda_{H S}|$ and $10^{10} \alpha_S^2$ using the same parameters as in figure~\ref{omega-lafs}. It is observed that the peak of scalar DM relic density (shown in figure~\ref{omega-lafs} (right)) aligns with the minimum of $ \alpha_S^2 $ (as seen in figure~\ref{lafsii}), confirming expectations.  

\section{Direct detection} \label{sec4}
The direct detection can identify DM particles, provided there is a chance for a collision between nucleons and DM particles. However, to date, no such events have been observed in direct detection experiments. In this section, we will calculate the spin-independent DM-nucleon cross section.

\begin{figure} [!htb] 
\centerline{\hspace{0cm}\epsfig{figure=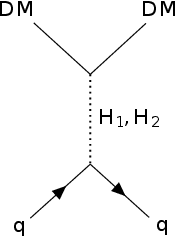,width=2.9cm}}
\caption{Relevant Feynman diagram for direct detection.} \label{FeynmanDD}
\end{figure}

In our model, all DM components interact with quarks via the exchange of $H_1$ or $H_2$ within the Higgs portal framework (see figure~\ref{FeynmanDD}), leading to a spin-independent cross section for DM-nucleon interactions. The energy scale for DM scattering on nuclei is around 1 GeV, with the momentum exchange between the DM and the nucleon being quite low. Additionally, the slow velocity of the DM allows us to analyze this process using a nonrelativistic approximation. As a result, the interactions of DM with nucleons can be captured by effective four-field interactions between the DM and the SM quarks. When we integrate out the scalar mediators, the resulting low-energy effective interaction takes the form of a 5-dimensional (6-dimensional) interaction between scalar and vector DM (spinor DM) and the quarks,

\begin{align} \label{lagr}
\mathcal{L}_{S-q} &= \alpha_S S^2 \sum_q m_q \overline{q} q , \nonumber \\
\mathcal{L}_{\psi-q} &= \alpha_\psi \sum_{a=1}^2 \bar\psi^a \psi^a \sum_q m_q \overline{q} q , \nonumber \\
\mathcal{L}_{V-q} &=  \alpha_V V_{\mu}V^{\mu} \sum_q m_q \overline{q} q,
\end{align}
where
\begin{align} \label{couplings}
& \alpha_S= \frac{\lambda_{HS}}{2}(\frac{\cos^2 \alpha}{M_{H_1}^2}+\frac{\sin^2 \alpha}{M_{H_2}^2})- \frac{\lambda_{\phi S} \cos^2 \alpha}{2} (\frac{1}{M_{H_1}^2}-\frac{1}{M_{H_2}^2}),   \nonumber \\
& \alpha_{\psi}=-\frac{g_{\phi} \sin \alpha \cos \alpha}{\sqrt{2}\nu_{1}}  (\frac{1}{M_{H_1}^2}-\frac{1}{M_{H_2}^2}), \nonumber \\
& \alpha_V= g_{V}^2 \cos^2 \alpha (\frac{1}{M_{H_1}^2}-\frac{1}{M_{H_2}^2}).
\end{align}
This allows for the derivation of effective interactions between the DM particles and a nucleon, resulting in the spin-independent DM-nucleon cross sections for scalar, spinor, and vector DM \cite{Kanemura:2010sh}:
\begin{align} \label{CSequations}
& \sigma_S= \alpha_{S}^2 \frac{M_{N}^4}{\pi (M_N + M_S)^2}f_{N}^2,  \nonumber \\
& \sigma_{\psi}= \alpha_{\psi}^2 \frac{M_{N}^4 M_{\psi}^2}{\pi (M_N + M_{\psi})^2}f_{N}^2,  \nonumber \\
& \sigma_V= \alpha_{V}^2 \frac{M_{N}^4}{\pi (M_N + M_V)^2}f_{N}^2,
\end{align}
where \( M_N \) represents the mass of the nucleon and \( f_N = 0.3 \) characterizes the coupling between the Higgs and nucleons. We have verified (\ref{CSequations}) once again using {\tt micrOMEGAs}~\cite{Alguero:2023zol}.

\begin{figure}[!htb] 
\begin{center}
\centerline{\hspace{0cm}\epsfig{figure=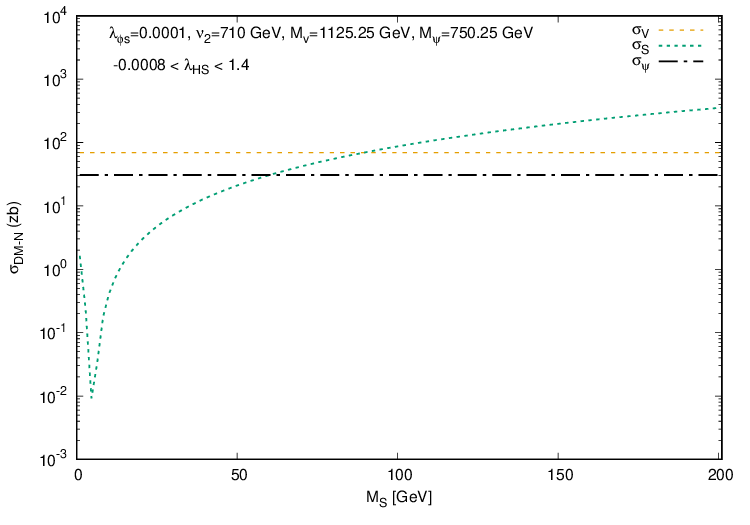,width=7.5cm}}
\centerline{\vspace{-0.7cm}}
\caption{DM-nucleon cross section as a function of scalar DM mass.} \label{scalarCS}
\end{center} 
\end{figure}

\begin{figure}[!htb] 
\begin{center}
\centerline{\hspace{0cm}\epsfig{figure=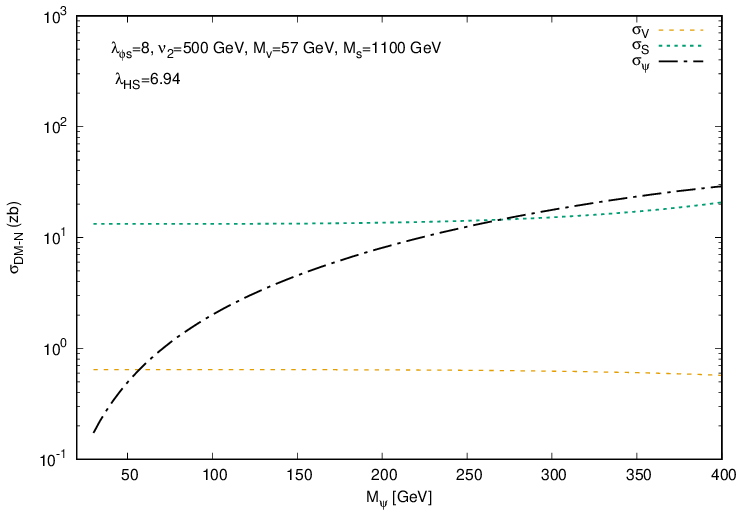,width=7.5cm}}
\centerline{\vspace{-0.7cm}}
\caption{DM-nucleon cross section as a function of spinor DM mass.} \label{spinorCS}
\end{center} 
\end{figure}

\begin{figure}[!htb] 
\begin{center}
\centerline{\hspace{0cm}\epsfig{figure=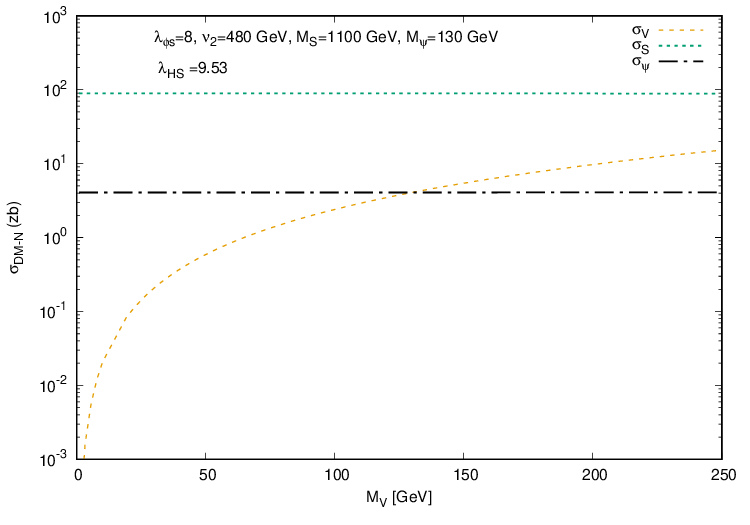,width=7.5cm}}
\centerline{\vspace{-0.7cm}}
\caption{DM-nucleon cross section as function of vector DM mass.} \label{vectorCS}
\end{center} 
\end{figure}

\begin{figure}[!htb] 
\begin{center}
\centerline{\hspace{0cm}\epsfig{figure=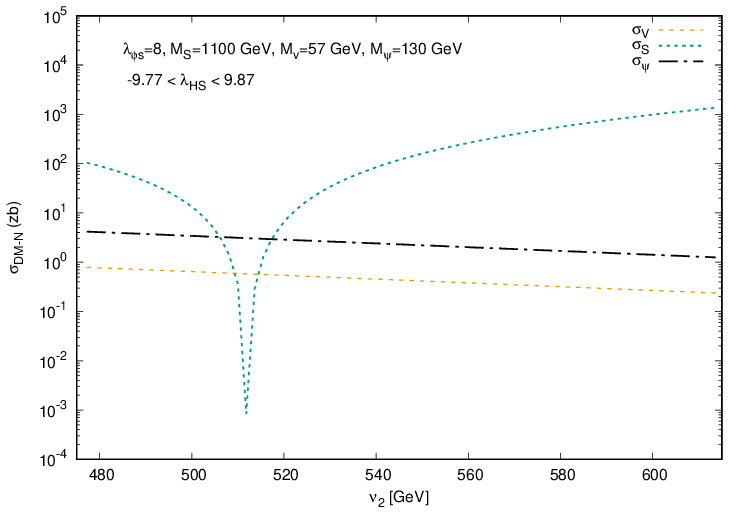,width=7.5cm}}
\centerline{\vspace{-0.7cm}}
\caption{DM-nucleon cross section as a function of $\nu_2$.} \label{nu2CS}
\end{center} 
\end{figure}

\begin{figure}[!htb] 
\begin{center}
\centerline{\hspace{0cm}\epsfig{figure=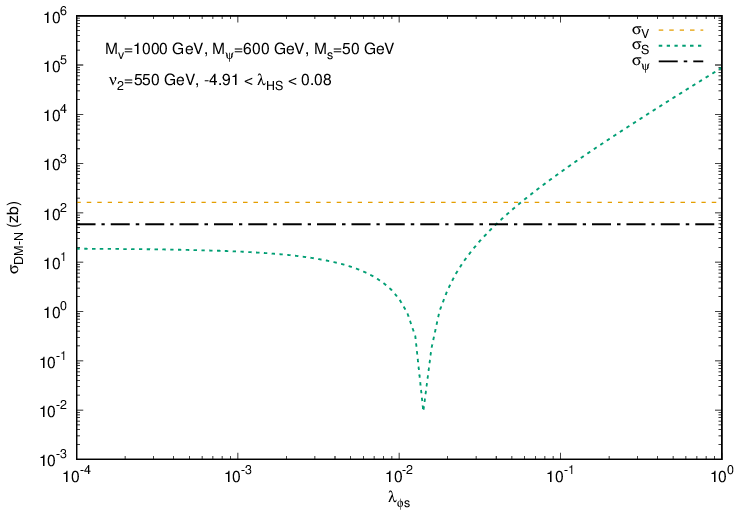,width=7.5cm}}
\centerline{\vspace{-0.7cm}}
\caption{DM-nucleon cross section as a function of $\lambda_{\phi S}$.} \label{lafsCS}
\end{center} 
\end{figure}

The DM-nucleon cross section for all DM components is presented in figures \ref{scalarCS} to \ref{lafsCS} as a function of the free parameters.
In figure \ref{scalarCS}, the lowest value of $\sigma_S$ occurs when $\lambda_{HS}$ is zero, resulting in a minimum in $\alpha_s$ near $M_S \approx 5$ GeV. Since $M_{H_2}$ (\ref{MH2}) remains relatively unchanged despite an increase in $M_S$ for the selected parameters, both $\sigma_{\psi}$ and $\sigma_{V}$ remain constant across these parameters.
In figure \ref{spinorCS}, the coupling constant $g_{\phi}$ increases as $M_{\psi}$ rises (see (\ref{constrins})) which in turn causes an increase in $\alpha_{\psi}$ (\ref{couplings}) and leads to an upward trend in $\sigma_{\psi}$. Similarly, in figure \ref{vectorCS}, since $g_V$ depends on $M_{V}$, we can see that $\sigma_V$ also increases with higher values of $M_{V}$.
Figure~\ref{nu2CS} illustrates the DM-nucleon cross section for all DM components as a function of $ \nu_2 $, while keeping other free parameters constant. As $ \nu_2 $ changes, it affects both $ \alpha $ and $ M_{H_2} $ in equation (\ref{couplings}), and changes the DM-nucleon cross sections. Note that $\lambda_{H S}$ approaches zero at approximately $\nu_2$=550~GeV; however, the lowest value of $\sigma_s$ occurs at $\nu_2\simeq510$~GeV, which corresponds to the minimum of $ \alpha_S^2 $. Moreover, we notice that as \( \nu_2 \) rises, both \( \sigma_{\psi} \) and \( \sigma_{V} \) decrease because the increase in \( \nu_2 \) leads to a reduction in \( \alpha_V^2 \) and \( \alpha_{\psi}^2 \).
In figure~\ref{lafsCS}, by fixing the values of $M_S$, $M_V$, $M_{\psi}$, and $\nu_2$, many of the key parameters in (\ref{couplings}), such as $\alpha$, $M_{H_2}$, $g_{\phi}$, and $g_V$, will also be determined. Consequently, only $\lambda_{\phi S}$ and $\lambda_{HS}$ will vary in relation to one another. As a result, it is anticipated that $\sigma_S$ will change with variations in $\lambda_{\phi S}$, while $\sigma_V$ and $\sigma_{\psi}$ will remain unaffected. Notably, around $\lambda_{\phi S} = 0.014$, we find that $\alpha_S^2 \simeq 0$ (see figure \ref{lafsii}), indicating a minimum in the scalar DM-nucleon cross section.

Current constraints on DM-nucleon spin-independent interactions are derived from cutting-edge experiments, including XENONnT~\cite{XENON:2023cxc} and PandaX-4T ~\cite{PandaX-4T:2021bab}.
Ultimately, the DARWIN experiment \cite{DARWIN:2016hyl} aims to achieve sensitivity close to the irreducible background caused by the scattering of SM neutrinos with nucleons, known as the neutrino floor \cite{Billard:2013qya}, positioning it as the premier detector for DM. In the next section, we analyze the constraints on our model imposed by the findings from the PandaX-4T experiment \cite{PandaX-4T:2021bab}, which established an upper limit on the spin-independent WIMP-nucleon cross section, with the lowest exclusion occurring at $M_{DM} = 40$ GeV:
\begin{equation} \label{pan}
\text{PandaX-4T}: \quad \sigma_{\text{DM-N}} \lesssim 3.8 \times 10^{-47} \, \text{cm}^2 .
\end{equation}
It is crucial to emphasize that the previous constraint is based on the assumption that the local DM density is derived exclusively from a single DM species. However, in the context of our scenario, this is not the case. Instead, various types of DM, including scalar, spinor, and vector particles, all contribute to the overall local DM density. This multiplicity of contributions demands a more subtle understanding of the DM composition in our model. To quantify the relative importance of each type of DM present, we will define DM fractions as follows:
\begin{equation} \label{DMfraction}
\xi_{S}=\frac{\Omega_{S}}{\Omega_{DM}}, \;\;\; \xi_{\psi}=\frac{\Omega_{\psi}}{\Omega_{DM}}, \;\;\; \xi_{V}=\frac{\Omega_{V}}{\Omega_{DM}}.
\end{equation}
where $ \xi_{S}+\xi_{\psi}+\xi_{V}=1. $
Assuming that the share of each DM candidate in the local DM density mirrors their contribution to the relic density, many researchers have imposed constraints on the rescaled DM-nucleon cross sections, namely $ \xi_{S} \sigma_S $, $ \xi_{\psi} \sigma_{\psi} $, and $ \xi_{V} \sigma_V $, based on experimental findings. Nevertheless, all DM candidates yield observable signatures, meaning that these signatures must be considered together. In cases of large DM mass, where the DM energy significantly exceeds the detector's threshold energy, the statistical combination is straightforward, and for $ M \gtrsim 40 $ GeV the resulting direct detection constraint is expressed as \cite{Hur:2007ur}
\begin{equation} \label{pan1}
\frac{\sigma}{\text{M}_{\text{DM}}} \equiv \xi_S \frac{\sigma_s}{M_S} + \xi_{\psi} \frac{\sigma_{\psi}}{M_{\psi}} + \xi_V \frac{\sigma_v}{M_V} \lesssim \frac{\sigma}{M} \bigg\rvert_{\text{PandaX-4T}} \simeq 0.0005 \, \frac{\text{zb}}{\text{GeV}}.
\end{equation}

Finally, it is important to highlight that, within the framework of our model, the constraints imposed by indirect detection do not match the competitiveness of those derived from direct detection methods. This discrepancy in efficacy means that indirect detection strategies provide less robust limits on the parameters we are investigating. Consequently, we will refrain from exploring this aspect in detail in this discussion, as our primary focus in the next section will remain on the more significant findings and implications associated with DM relic density and direct detection.

\section{Results} \label{sec5}
In this section, we begin by imposing constraints on the five free parameters of our model, as outlined in (\ref{freeparameters}). These constraints are based on empirical data from the Planck satellite mission \cite{Planck:2018vyg}, particularly focusing on the DM relic density in equation (\ref{0.12}). Following this initial analysis, we evaluate the constrained parameters for compatibility with results from direct detection experiments by the XENONnT collaboration \cite{XENON:2023cxc} and the PandaX-4T experiment \cite{PandaX-4T:2021bab}.

\begin{figure}[!htb] 
\begin{center}
\centerline{\hspace{0cm}\epsfig{figure=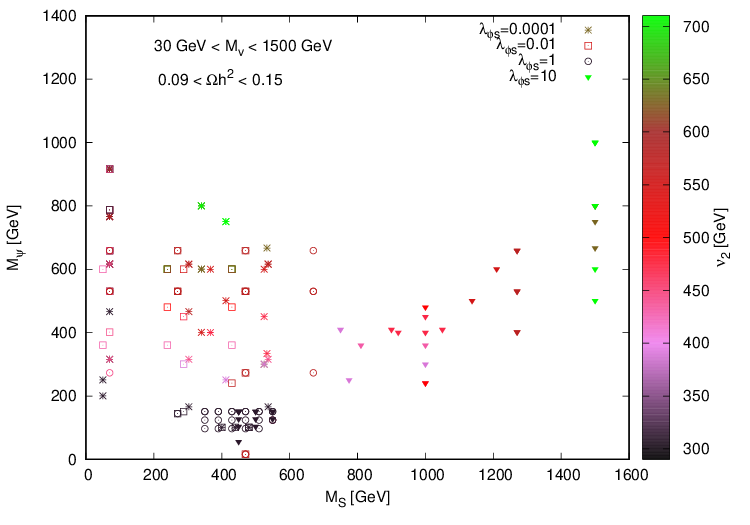,width=6.5cm}\hspace{0cm}\epsfig{figure=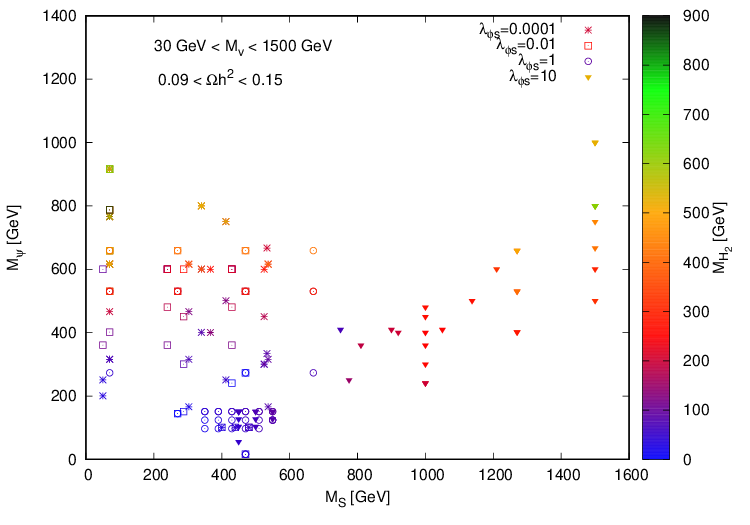,width=6.5cm}}
\centerline{\vspace{-0.7cm}}

\centerline{\hspace{0cm}\epsfig{figure=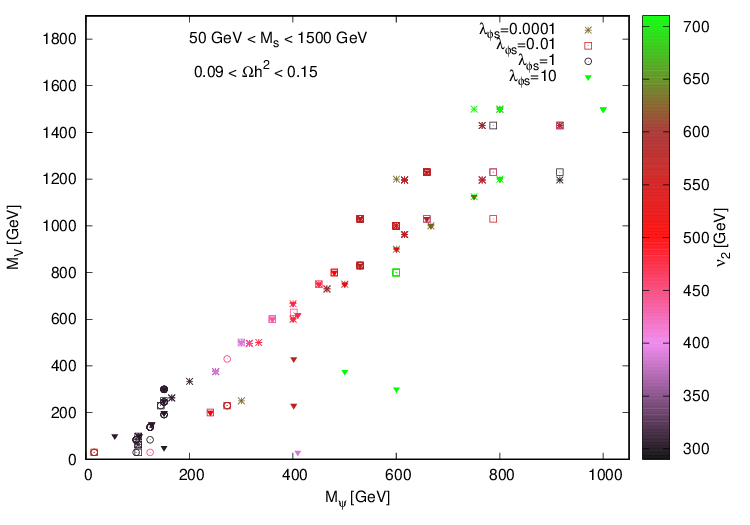,width=6.5cm}\hspace{0cm}\epsfig{figure=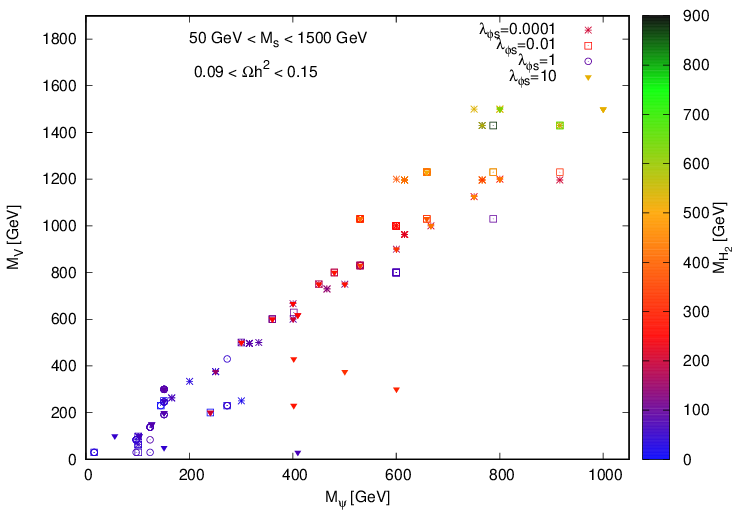,width=6.5cm}}
\centerline{\vspace{-0.7cm}}

\centerline{\hspace{0cm}\epsfig{figure=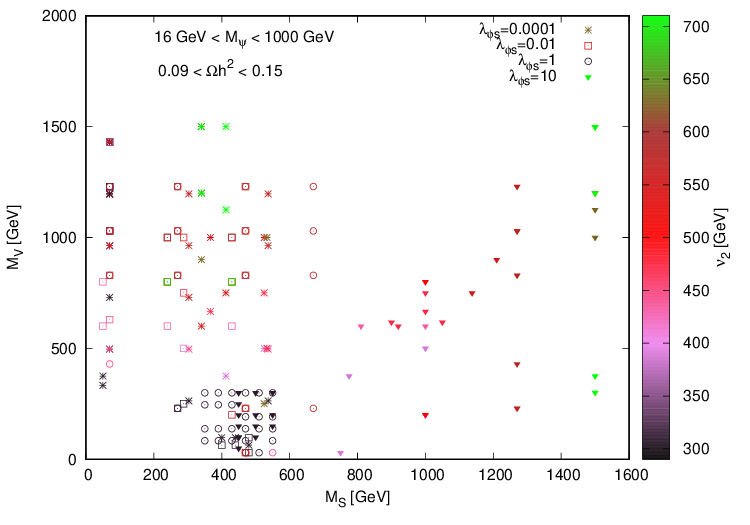,width=6.5cm}\hspace{0cm}\epsfig{figure=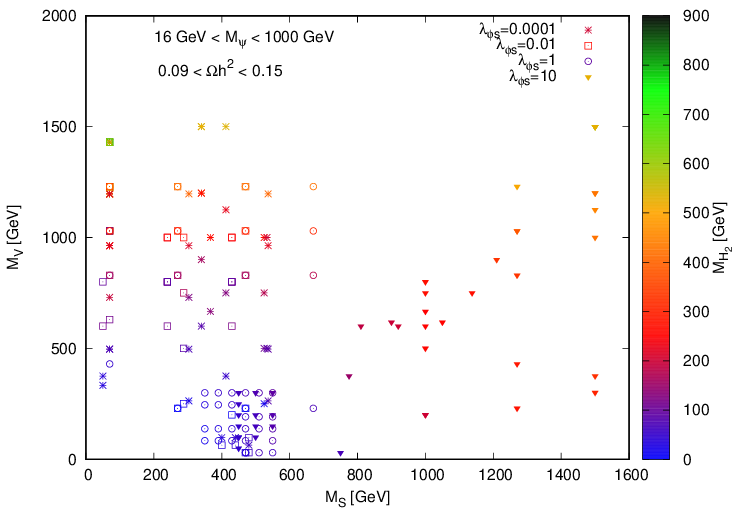,width=6.5cm}}
\centerline{\vspace{-0.7cm}}

\caption{The permissible parameter space is shaped by the constraint established by the DM relic density measurements reported by the Planck collaboration. These specified regions offer a framework for comprehending how different parameters work together to produce results that align with existing observational data.} \label{Mdm}
\end{center} 
\end{figure}

In our approach, we have selected four distinct values for the coupling parameter \( \lambda_{\phi S} \) ($ \lambda_{\phi S}=10^{-4}, 10^{-2}, 1, 10 $). Following this, we systematically explored the ranges of the other free parameters within well-defined domains:

\begin{center}
50 GeV \( < M_{S} < \) 1500 GeV, \\ 
16 GeV \( < M_{\psi} < \) 1000 GeV, \\
30 GeV \( < M_{V} < \) 1500 GeV, \\ 
and 290 GeV \( < \nu_{2} < \) 710 GeV.
\end{center}

In addition to DM relic density and direct detection boundaries, we have taken into account perturbativity, as illustrated in (\ref{4pi}), along with the requirement for vacuum stability, which necessitates that \( M_{H_2}^2 > 0 \) as presented in equation (\ref{MH2}). These theoretical constraints play a crucial role in ensuring the validity of our model. 

The resultant parameter space that satisfies the condition \( 0.09 < \Omega h^2 < 0.15 \) has been depicted in figure~\ref{Mdm}. This visualization elucidates the interplay between the various parameters and highlights the regions of interest for DM relic density in our analysis.

\begin{figure} [!htb] 
\centerline{\hspace{0cm}\epsfig{figure=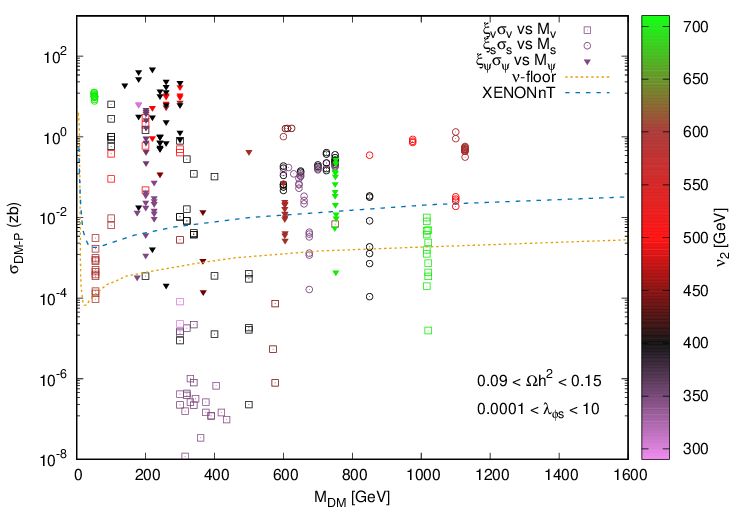,width=6.5cm}\hspace{0cm}\epsfig{figure=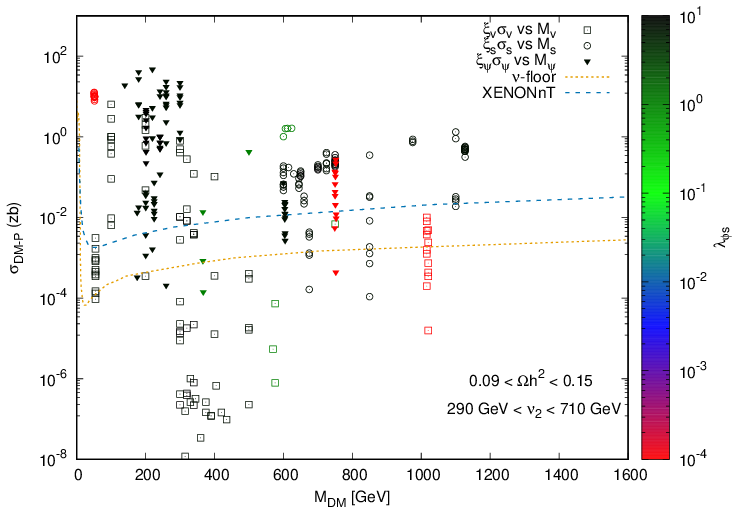,width=6.5cm}}
\caption{The rescaled DM-nucleon cross section within a parameter space constrained by DM relic density. The plots show the values of the independent parameters \( \nu_2 \) and \( \lambda_{\phi S} \) via a color bar. The inclusion of the neutrino floor highlights the sensitivity limit in direct detection experiments, emphasizing the challenges in detecting DM signals.} \label{xenon}
\end{figure}

In figure \ref{xenon}, we present an illustration of all rescaled DM-nucleon elastic scattering scenarios within a specific parameter space that has already been constrained by the relic density limit established by Planck data. To enhance clarity and understanding, we have included two distinct plots that represent the values of the two independent parameters, namely $ \nu_2 $ and $ \lambda_{\phi S} $, which are indicated through a color bar for ease of interpretation. This depiction clearly demonstrates that there exists a viable parameter space that not only aligns with the observational requirements derived from the Planck data but also adheres to the DM-Nucleon upper bound established by the XENONnT experiment. Additionally, we have included the neutrino floor in the figure, which represents a fundamental limit on the sensitivity of direct detection experiments due to the background noise caused by neutrinos produced from various astrophysical sources, primarily from the Sun and cosmic neutrinos. This crucial element emphasizes the challenges faced in detecting DM, as it indicates the lowest threshold for signal detection amidst the background noise created by neutrinos.

\begin{figure} [!htb] 
\centerline{\hspace{0cm}\epsfig{figure=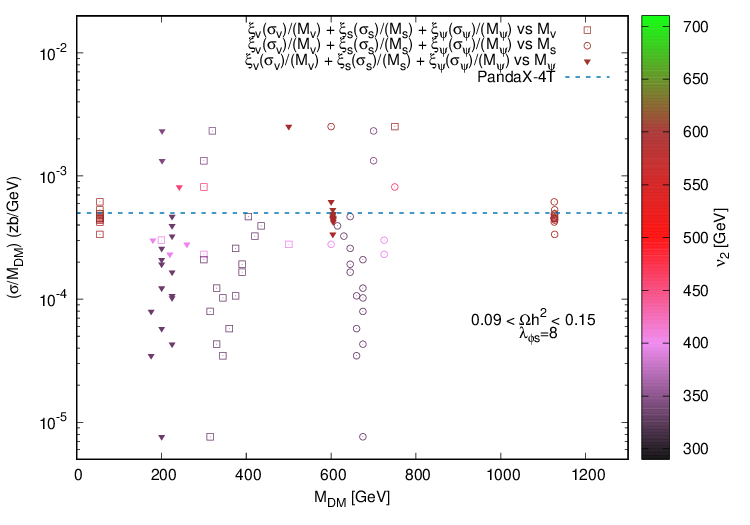,width=7.5cm}}
\caption{This figure illustrates the stringent criteria for direct detection of DM as defined in equation (\ref{pan1}). The plot displays the ratio \( \sigma / \text{M}_{\text{DM}} \) against the masses of the DM components for $ \lambda_{\phi S} = 8 $.} \label{panda}
\end{figure}

In figure \ref{panda}, we delve into a more stringent set of criteria for direct detection, specifically focusing on the criterion outlined in equation (\ref{pan1}), which applies to all DM components. In this analysis, we illustrate the relationship between the ratio \( \sigma / \text{M}_{\text{DM}} \), as defined in equation (\ref{pan1}), and the masses of the various DM components. 

It is important to note that we have excluded values for \( \lambda_{\phi S} = 10^{-4} \) and \( 10^{-2} \) from our presentation. This exclusion is due to the fact that the corresponding \( \sigma / \text{M}_{\text{DM}} \) values for these parameters significantly exceed the upper limit established by the PandaX-4T experiment. Consequently, our focus is narrowed to only those data points that are either close to or fall below the threshold of \( \sigma / \text{M}_{\text{DM}} = 5 \times 10^{-4} \) GeV/zb.

To enhance clarity and understanding, $ \nu_2 $ is indicated through a color bar, allowing for an intuitive grasp of the relationships and variations present in the data. This comprehensive approach not only highlights the pertinent values but also facilitates a deeper understanding of the dynamics at play in the context of DM direct detection. 

\section{Conclusion} \label{sec6}
In this study, we have explored a novel model of multi-component DM that extends beyond the SM by introducing five new fields: a scalar field ($ S $), two spinor fields ($ \psi^{1,2} $), a vector field ($ V_{\mu} $), and an additional scalar field ($ \phi $) that interacts with the SM through a dark $ U_D(1) $ gauge symmetry. This framework not only addresses the longstanding issues of DM composition but also tackles the hierarchy problem and vacuum instability inherent in the SM.

Our analysis reveals that all DM components—scalar, spinor, and vector—each characterized by distinct spins, can collectively contribute to the observed DM relic density. By employing Boltzmann equations, we calculated the DM abundances, taking into account various processes such as annihilations, semi-annihilations, and conversions. The results indicate that the model is viable across a broad range of DM masses, with parameter scans aligning with current experimental constraints, particularly from direct detection experiments like XENONnT and PandaX-4T.

The model maintains classical scale invariance, while electroweak symmetry breaking occurs due to loop effects, allowing for the dynamic generation of particle masses through the Coleman-Weinberg mechanism. This approach provides a robust framework for understanding the interplay between different DM components and their interactions with the SM.

Our findings emphasize the importance of multi-component DM models in addressing the complexities of DM physics. The constraints imposed by the observed DM relic density, as reported by the Planck collaboration, significantly limit the parameter space of our model, guiding us toward feasible regions that are consistent with experimental data. The analysis of direct detection cross sections further highlights the potential for future experiments to probe the nature of DM, especially as they approach the neutrino floor, which represents the ultimate sensitivity limit for detecting DM interactions.

In conclusion, our study not only enhances the understanding of DM but also opens avenues for future research in exploring the implications of new physics beyond the SM. The comprehensive framework established here could lead to promising signals in upcoming direct detection experiments, contributing to the ongoing quest to decipher the nature of the Universe's unseen constituents. As we continue to refine our model and incorporate new experimental data, we remain optimistic about the prospects for discovering the fundamental properties of dark matter and its role in the cosmos.

\section*{Acknowledgments}
We would like to express our gratitude to the referee for their insightful feedback. Their observation regarding the anomaly in the initial version of our model, which included only a single spinor field, was instrumental in guiding us to refine and improve the work. This constructive critique has significantly strengthened the robustness of our model.

\providecommand{\href}[2]{#2}\begingroup\raggedright\endgroup

\end{document}